\documentclass[aps,prc,twocolumn,floatfix,nofootinbib,showpacs,superscriptaddress]{revtex4-1}

\usepackage{amsmath,amsfonts,amssymb,bm}
\usepackage{graphicx}
\usepackage{color}
\usepackage{pzccal}
\usepackage{soul}

\begin{document}

\newcommand{\dslash}[1]{#1 \hspace{-1.2ex}\slash}

\title{Practical corollaries of transverse Ward-Green-Takahashi identities}

\author{Si-xue Qin}
\affiliation{Department of Physics, Center for High Energy Physics and State Key Laboratory of Nuclear Physics and Technology, Peking University, Beijing 100871, China}
\affiliation{Institut f\"ur Theoretische Physik, Johann Wolfgang Goethe University, Max-von-Laue-Str.\,1, D-60438 Frankfurt am Main, Germany}

\author{Lei Chang}
\affiliation{Institut f\"ur Kernphysik, Forschungszentrum J\"ulich, D-52425 J\"ulich, Germany}

\author{Yu-xin Liu}
\email[Corresponding author: ]{yxliu@pku.edu.cn}
\affiliation{Department of Physics, Center for High Energy Physics and State Key Laboratory of Nuclear Physics and Technology, Peking University, Beijing 100871, China}

\author{Craig D. Roberts}
\email[Corresponding author: ]{cdroberts@anl.gov}
\affiliation{Physics Division, Argonne National Laboratory, Argonne, Illinois 60439, USA}
\affiliation{Department of Physics, Illinois Institute of
Technology, Chicago, Illinois 60616-3793, USA}

\author{Sebastian M.~Schmidt}
\affiliation{Institute for Advanced Simulation, Forschungszentrum J\"ulich and JARA, D-52425 J\"ulich, Germany}

\date{14 February 2013}

\begin{abstract}
The gauge principle is fundamental in formulating the Standard Model.  Fermion--gauge-boson couplings are the inescapable consequence and the primary determining factor for observable phenomena.  Vertices describing such couplings are simple in perturbation theory and yet the existence of strong-interaction bound-states guarantees that many phenomena within the Model are nonperturbative.  It is therefore crucial to understand how dynamics dresses the vertices and thereby fundamentally alters the appearance of fermion--gauge-boson interactions.  We consider the coupling of a dressed-fermion to an Abelian gauge boson, and describe a unified treatment and solution of the familiar longitudinal Ward-Green-Takahashi identity and its less well known transverse counterparts.  Novel consequences for the dressed-fermion--gauge-boson vertex are exposed.
\end{abstract}


\maketitle

\begin{figure}[t]
\centerline{%
\includegraphics[clip,width=0.45\linewidth]{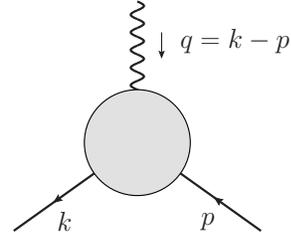}}
\caption{\label{figvertex} Dressed-fermion--gauge-boson vertex, $\Gamma_\mu(k,p)$, with the momentum flow indicated.}
\end{figure}

\hspace*{-\parindent}\textbf{Introduction}.  Identities of the Ward-Green-Takahashi (WGT) type \cite{Ward:1950xp,Green:1953te,Takahashi:1957xn} have long been known and used in gauge theories.  The widely familiar forms provide constraints on the longitudinal part of $n$-point Schwinger functions; i.e., propagators and vertices.  For example, in an Abelian gauge theory the dressed-fermion--gauge-boson vertex, $\Gamma_\mu(k,p)$ in Fig.\,\ref{figvertex}, satisfies
\begin{equation}
\label{eqWGTI}
q_\mu i \Gamma_\mu(k,p) = S^{-1}(k) - S^{-1}(p)\,,
\end{equation}
where the dressed-fermion propagator may be written 
\begin{equation}
S(p) = 
1/[i\gamma\cdot p A(p^2) + B(p^2)]\,.
\end{equation}
%
Equation~\eqref{eqWGTI} is a nonperturbative consequence of gauge invariance in an Abelian theory and, following Ref.\,\cite{Ball:1980ay}, it has been used extensively in the construction of models for the dressed-fermion--gauge-boson vertex.  (Renormalisation does not affect the form of the identities we consider, so we do not explicitly refer to it.  A Euclidean metric is used herein:  $\{\gamma_\mu,\gamma_\nu\} = 2\delta_{\mu\nu}$; $\gamma_\mu^\dagger = \gamma_\mu$; $\gamma_5= \gamma_4\gamma_1\gamma_2\gamma_3$, tr$[\gamma_5\gamma_\mu\gamma_\nu\gamma_\rho\gamma_\sigma]=-4 \epsilon_{\mu\nu\rho\sigma}$; $\sigma_{\mu\nu}=(i/2)[\gamma_\mu,\gamma_\nu]$; $a \cdot b = \sum_{i=1}^4 a_i b_i$; and $Q_\mu$ spacelike $\Rightarrow$ $Q^2>0$.)

It is natural to question the need for vertex \emph{Ans\"atze}, given that QED has been tested perturbatively to remarkable precision through comparison with experiment \cite{Mohr:2012tt}.  However, the context for utility here is not the QED of interactions between leptons and photons.  Instead, one has in mind theories in which the dressed-fermion propagator is strongly modified from its perturbative behaviour and hence so, too, is the vertex.

An obvious example is QCD, in which the dressed-quark two-point function is described by two momentum-dependent functions: a wave-function renormalisation, $Z(p^2)=1/A(p^2)$, and mass function, $M(p^2)=B(p^2)/A(p^2)$, both of which are strongly modified from their perturbative forms for $p^2 \lesssim (5\,\Lambda_{\rm QCD})^2$ \cite{Bhagwat:2006tu,Kamleh:2007ud}.  This dressing is associated with dynamical chiral symmetry breaking (DCSB), a particularly striking feature of the Standard Model that plays an important role in formation of the visible matter in the Universe \cite{national2012Nuclear}.  In a massless theory; i.e., in QCD's chiral limit in the absence of DCSB, $M(p^2)\equiv 0$.  (Notably, the gluon also acquires a momentum-dependent mass function \cite{Boucaud:2011ug,LlanesEstrada:2012my}, $m_G(k^2)$, which is large at infrared momenta \cite{Bowman:2004jm,Cucchieri:2011ig,Wilson:2012np,Aguilar:2012rz,Ayala:2012pb,Binosi:2012sj,Strauss:2012dg,Oliveira:2012uy}: $m_G(0) \simeq 0.5\,$GeV$\,\gtrsim M(0)$.)

The electromagnetic vertex associated with such a quark must differ markedly from the perturbative form at infrared momenta.  This is obvious from Eq.\,\eqref{eqWGTI}, since
\begin{equation}
i q_\mu \gamma_\mu \neq i \gamma\cdot k A(k^2) + B(k^2) - i \gamma\cdot p A(p^2) - B(p^2)\,;
\end{equation}
and the mismatch is largest within that domain upon which $A$, $B$ differ conspicuously from their perturbative forms.  The importance of this and kindred dressing to the reliable computation of observables involving composite systems was exposed in Refs.\,\cite{Frank:1994mf,Roberts:1994hh,Munczek:1994zz,Bender:1996bb}.  The most sophisticated \emph{Ans\"atze} currently available are detailed in Refs.\,\cite{Kizilersu:2009kg,Bashir:2011dp} but they follow upon a great deal of work, which may be traced from Ref.\,\cite{Ball:1980ay}.

\smallskip

\hspace*{-\parindent}{\textbf{Transverse Identities}}.  The transverse WGT identities \cite{Takahashi:1985yz,Kondo:1996xn,He:2000we,He:2006my,He:2007zza} are less familiar and, \emph{prima facie}, less useful, too.  In (colour-singlet) vector and axial-vector channels connected with a fermion of mass $m$, these identities read $(t=k+p)$:
\begin{eqnarray}
\nonumber
\lefteqn{q_\mu \Gamma_\nu(k,p)-q_\nu \Gamma_\mu(k,p) =
S^{-1}(p)\sigma_{\mu\nu} + \sigma_{\mu\nu} S^{-1}(k) }\\
&& + 2 i m \Gamma_{\mu\nu}(k,p) + t_\lambda \varepsilon_{\lambda\mu\nu\rho} \Gamma^A_{\rho}(k,p) + A^{V}_{\mu\nu}(k,p)\,,
\label{eqTWGTI}\\
%
\nonumber
\lefteqn{q_\mu \Gamma^A_{\nu}(k,p) - q_\nu \Gamma^A_{\mu}(k,p)=
S^{-1}(p)\sigma^5_{\mu\nu} -\sigma^5_{\mu\nu}S^{-1}(k)}\\
&& + t_\lambda\varepsilon_{\lambda\mu\nu\rho} \Gamma_\rho(k,p) + V^{A}_{\mu\nu}(k,p)\,,
\label{eqTAWGTI}
\end{eqnarray}
where $\sigma^5_{\mu\nu} = \gamma_5\sigma_{\mu\nu}$ and $\Gamma_{\mu\nu}(k,p)$ is an inhomogeneous tensor vertex.  Whereas the longitudinal WGT identity expresses properties of the divergence of the vertex, the transverse identities relate to its \emph{curl} (as Faraday's law of induction involves an electric field).
The last two terms in Eq.\,\eqref{eqTWGTI} arise in computing the momentum space expression of a nonlocal axial-vector vertex, whose definition involves a gauge-field-dependent line integral \cite{He:2007zza}; and the last two terms in Eq.\,\eqref{eqTAWGTI} arise from similar manipulations of an analogous nonlocal vector vertex.  Note that, like Eq.\,\eqref{eqWGTI}, the transverse identities are valid in any covariant gauge, which is the class we focus upon, and do not explicitly display dependence on the gauge-fixing parameter.  (Whilst an anomaly term can be included in Eq.\,\eqref{eqTAWGTI} \cite{He:2002jg}, we concern ourselves with flavoured vertices; i.e., those free from such amendment.)

It is the presence of the unfamiliar quantities $A^V_{\mu\nu}(k,p)$, $V^A_{\mu\nu}(k,p)$ in the transverse identities that lends them an appearance of impracticality, since even at one-loop order the expressions for $A^V_{\mu\nu}(k,p)$, $V^A_{\mu\nu}(k,p)$ are complicated \cite{He:2002jg,Pennington:2005mw,He:2006ce} and, moreover, they lead to a coupling between the vector and axial-vector identities.  We cannot overcome the complexity but something can be done about the induced coupling between the identities.

Before doing so, it is worthwhile explaining that, in general, twelve independent tensor structures are required to specify a fermion--vector-boson vertex: given the matrix-valued vectors $\gamma_\mu$, $k_\mu\mathbf{I}_{\rm D}$, $p_\mu\mathbf{I}_{\rm D}$, where $\mathbf{I}_{\rm D}$ is the $4\times 4$ identity in Dirac space, one can construct twelve independent quantities that behave as a vector under Poincar\'e transformations.

We make our conventions explicit by writing
{\allowdisplaybreaks
\begin{eqnarray}
\Gamma_\mu(k,p) &=& \Gamma_\mu^{L}(k,p) + \Gamma_\mu^T(k,p)\,,\\
\label{eqGammaL}
\Gamma_\mu^{L}(k,p) &=& \sum_{j=1}^4 \lambda_j(k,p) \, L^j_\mu(k,p)\,, \\
\label{eqGammaT}
\Gamma_\mu^{T}(k,p) &=& \sum_{j=1}^8 \tau_j(k,p) \, T^j_\mu(k,p)\,,
\end{eqnarray}}
\hspace*{-0.4\parindent}where the matrix-valued tensors $\{L^j_\mu(k,p),j=1,\ldots,4\}$ and $\{T^j_\mu(k,p),j=1,\ldots,8\}$ are given, respectively, in Eqs.\,\eqref{eqlambdaG} and \eqref{TsEuclidean}.  Following inspection of Eqs.\,\eqref{TsEuclidean}, it becomes clear that $q_\mu \Gamma_\mu^{T}(k,p) \equiv 0$.

Now consider the matrix-valued tensors
\begin{equation}
T^1_{\mu\nu} =
%
\frac{1}{2} \,  \varepsilon_{\alpha\mu\nu\beta} t_\alpha q_\beta \mathbf{I}_{\rm D}\,,\;
T^2_{\mu\nu} =
\frac{1}{2} \, \varepsilon_{\alpha\mu\nu\beta} \gamma_\alpha q_\beta \,.
\end{equation}
Contracting the left-hand-side of Eq.\,\eqref{eqTAWGTI} with these tensors produces zero. Operating then with the right-hand-sides equated to zero, and using
\begin{eqnarray}
\nonumber
\lefteqn{T^1_{\mu\nu} t_\lambda\varepsilon_{\lambda\mu\nu\rho} \Gamma_\rho(k,p)}\\
&=& t^2 \, q\cdot \Gamma(k,p) - q\cdot t \, t\cdot \Gamma(k,p) ,\quad \\
%
\nonumber
\lefteqn{T^2_{\mu\nu}  t_\lambda\varepsilon_{\lambda\mu\nu\rho} \Gamma_\rho(k,p)}\\
&=& \gamma\cdot t \, q\cdot \Gamma(k,p) - q\cdot t \gamma \cdot \Gamma(k,p) \,,
\end{eqnarray}
one finds
\begin{eqnarray}
q\cdot t\, t \cdot \Gamma(k,p) &=& T^1_{\mu\nu} \left[S^{-1}(p)\sigma^5_{\mu\nu} - \sigma^5_{\mu\nu}S^{-1}(k)\right]  \nonumber\\
&& + \, t^2  q\cdot \Gamma(k,p) + T^1_{\mu\nu} V^{A}_{\mu\nu}(k,p) ,\quad
\label{eqtransWTI1}\\
%
q\cdot t\, \gamma \cdot \Gamma(k,p) &=& T^2_{\mu\nu} \left[S^{-1}(p)\sigma^5_{\mu\nu} - \sigma^5_{\mu\nu}S^{-1}(k)\right]\nonumber\\
&& + \, \gamma\cdot t \, q\cdot \Gamma(k,p) + T^2_{\mu\nu} V^{A}_{\mu\nu}(k,p). \quad
\label{eqtransWTI2}
\end{eqnarray}
This series of identities involves only the vector vertex.  We have thus uncoupled the equations.

It is worth remarking that the vector and axial-vector vertices together are required in order to understand electroweak interactions within the Standard Model.  One may obtain information about the axial-vector vertex through an analogous procedure involving Eq.\,\eqref{eqTWGTI}.

The complicated elements in Eqs.\,\eqref{eqtransWTI1}, \eqref{eqtransWTI2} are $T^{1,2}_{\mu\nu} V^{A}_{\mu\nu}(k,p)$.  Whilst these quantities cannot readily be computed, they are, nevertheless, merely matrix-valued scalar amplitudes and hence can be expressed succinctly:
\begin{eqnarray}
\nonumber
\lefteqn{iT^1_{\mu\nu} V^{A}_{\mu\nu}(k,p)= \mathbf{I}_{\rm D}\, X_1(k,p) + \gamma\cdot q\, X_2(k,p)}\\
&& \quad\quad + \gamma\cdot t\, X_3(k,p) + [\gamma\cdot q,\gamma\cdot t] X_4(k,p), \\
\nonumber
\lefteqn{iT^2_{\mu\nu} V^{A}_{\mu\nu}(k,p)= \mathbf{I}_{\rm D}\, X_5(k,p) + \gamma\cdot q\, X_6(k,p)}\\
&& \quad\quad + \gamma\cdot t\, X_7(k,p) + [\gamma\cdot q,\gamma\cdot t] X_8(k,p),
\end{eqnarray}
where $\{X_i,i=1,\ldots,8\}$ are scalar functions, which are undetermined until one has an \emph{Ansatz} or solution for the vector vertex.  We note that the mass-dimensions of $T^{1,2}_{\mu\nu} V^{A}_{\mu\nu}(k,p)$ are, respectively, three and two.

Although it might not be immediately obvious, these manipulations have achieved an important end.  They have brought us to a set of three matrix-valued equations for scalar-valued projections of $\Gamma_\mu(k,p)$; viz., Eqs.\,\eqref{eqWGTI}, \eqref{eqtransWTI1}, \eqref{eqtransWTI2}.  This amounts to a collection of twelve linearly-independent, coupled linear equations for twelve unknown scalar functions; and therefore a solution of these equations completely determines $\Gamma_\mu(k,p)$.

\smallskip

\hspace*{-\parindent}\textbf{Solution of the Coupled Identities}.  One may now use any reliable means to solve the system of coupled linear equations.  Irrespective of the presence and form of the functions $\{X_i,i=1,\ldots,8\}$, part of the complete solution has
\begin{equation}
\label{eqlambdaBC}
\begin{array}{ll}
\lambda_1(k,p) = \Sigma_A(k^2,p^2)\,, & \lambda_2(k,p) = \Delta_A(k^2,p^2)\,,\\
\lambda_3(k,p) = \Delta_B(k^2,p^2)\,, & \lambda_4(k,p) = 0\,,
\end{array}
\end{equation}
where
\begin{equation}
\begin{array}{l}
\Sigma_{\phi}(k^2,p^2) = \frac{1}{2}[\phi(k^2)+\phi(p^2)]\,, \\[1ex]
\Delta_{\phi}(k^2,p^2) = \frac{\phi(k^2)-\phi(p^2)}{k^2-p^2}\,.
\end{array}
\end{equation}
Namely, a necessary consequence of solving Eqs.\,\eqref{eqWGTI}, \eqref{eqtransWTI1}, \eqref{eqtransWTI2}, is the identification of $\Gamma_\mu^{L}(k,p)$ with the result derived in Ref.\,\cite{Ball:1980ay}; i.e, the Ball-Chiu \emph{Ansatz}.  The system of equations is linear, so the solution for $\Gamma^L_\mu(k,p)$ is \emph{unique}. Note that we made no attempt to impose a particular kinematic structure on the solution.  Irrespective of the tensor basis chosen, and we used a variety of forms, not just those in Eqs.\,\eqref{eqlambdaG}, \eqref{TsEuclidean}, this part of the solution is free of kinematic singularities.  The functional form of $\lambda_3(k,p)$ signals that the coupling of a dressed-fermion to a gauge boson is necessarily influenced heavily by DCSB.

The eight functions in Eq.\,\eqref{eqGammaT} are also completely specified.  Their form depends on $\{X_i,i=1,\ldots,8\}$; e.g.,
\begin{equation}
\tau_1(k,p) = \frac{1}{2} \frac{X_1(k,p)}{(k^2-p^2)((k\cdot p)^2-k^2 p^2)}\,.
\end{equation}
The expressions for $\{\tau_j,j=2,4,6,7\}$ are more complicated but, in common with $\tau_1$, they do not \emph{explicitly} involve the scalar functions ($A$, $B$) which define the dressed-fermion propagator.  It appears, therefore, that any and all effects of ($A$, $B$) in $\{\tau_j,j=1,2,4,6,7\}$ are only expressed implicitly through a solution of the vertex Bethe-Salpeter equation.

In contrast, the expressions for $\{\tau_j,j=3,5,8\}$ explicitly involve combinations of $A(k^2)$, $A(p^2)$, $B(k^2)$, $B(p^2)$ and $\{X_i,i=1,\ldots,8\}$.  If one supposes that $\{X_i\equiv 0,i=1,\ldots,8\}$, then simple results are obtained:
\begin{subequations}
\label{eqtau}
\begin{eqnarray}
2 \tau_3(k,p) & = & 
\Delta_A(k^2,p^2) \label{eqtau3} \,,\\
\tau_5(k,p) & = & - \Delta_B(k^2,p^2)\,, \label{eqtau5}\\
\tau_8(k,p) & = & - \Delta_A(k^2,p^2)\, .\label{eqtau8}
\end{eqnarray}
\end{subequations}

The following features of the transverse part of the solution to Eqs.\,\eqref{eqWGTI}, \eqref{eqtransWTI1}, \eqref{eqtransWTI2} are particularly noteworthy.

A $T^3_\mu(k,p)$ term generally appears in the solution and, with $\{X_i\equiv 0,i=1,\ldots,8\}$, its coefficient is $(1/2) \Delta_A(k^2,p^2)$, Eq.\,\eqref{eqtau3}.  The functional form is a prediction of the transverse WGT identity because, apart from our choice of tensor bases in Eqs.\,\eqref{eqlambdaG}, \eqref{TsEuclidean}, we implemented no other constraints.  Based upon considerations of multiplicative renormalisability and one-loop perturbation theory, a vertex \emph{Ansatz} was proposed in Ref.\,\cite{Bashir:2011dp}.  It involves a $T^3_\mu(k,p)$ term whose coefficient is $a_3\Delta_A(k^2,p^2)$, with $a_3+a_6 = 1/2$, where $a_6$ is associated with the $T^6_\mu(k,p)$ term in Eq.\,\eqref{eqGammaT}.  The agreement between the coefficients' functional forms is remarkable.  The choice $(a_3=0,a_6=1/2)$ produces the Curtis-Pennington \emph{Ansatz} \cite{Curtis:1990zs}.  The system of equations we have solved prefers the alternative $(a_3=1/2,a_6=0)$.  Corrections to Eq.\,\eqref{eqtau3} involve $\{X_i,i=2,3,5\}$.  They will depend on the gauge parameter and can affect the balance between $a_3$ and $a_6$ on that domain within which it is meaningful to think in these terms.  Curiously, then, the existence and strength of a Curtis-Pennington-like term in the vertex is determined by the nonlocal quantity $V^A_{\mu\nu}(k,p)$ in Eq.\,\eqref{eqTAWGTI}.

The solution contains an explicit anomalous magnetic moment term for the dressed-fermion; viz., a $T^5_\mu(k,p)$ term.  We find that its appearance is a straightforward consequence of Lagrangian-based symmetries but its necessary existence has been argued by other authors using very different reasoning \cite{Bicudo:1998qb,Kochelev:1996pv,Diakonov:2002fq,Chang:2010hb}.  With $\{X_i\equiv 0,i=1,\ldots,8\}$, the coefficient of $T^5_\mu(k,p)$ is ``$=-1\times \Delta_B(k^2,p^2)$;''   i.e., Eq.\,\eqref{eqtau5}.  We reiterate that the functional form is a prediction.  It signals  the intimate connection of this term with DCSB.  In Ref.\,\cite{Bashir:2011dp}, following a line of argument based upon multiplicative renormalisability and leading-order perturbation theory, a vertex \emph{Ansatz} was proposed in which the coefficient of this term is ``$-4/3 \times \Delta_B(k^2,p^2)$.''  The latter analysis was performed in Landau gauge whereas, herein, we have not needed to specify a gauge-parameter value.  The perfect agreement between the functional forms is striking and the near agreement between the coefficients is interesting.  Corrections to Eq.\,\eqref{eqtau5} involve $\{X_i,i=1,4,6\}$.  They will depend on the gauge parameter and can modify the coefficient on that domain within which it is meaningful to characterise the vertex \emph{Ansatz} in such a manner.  Thus, the strength of the explicit anomalous magnetic moment term in the vertex is finally determined by the nonlocal quantity $V^A_{\mu\nu}(k,p)$ in Eq.\,\eqref{eqTAWGTI}.

It was explained in Ref.\,\cite{Chang:2010hb} that in the presence of an explicit anomalous magnetic moment term, agreement with perturbation theory requires the appearance of $\tau_8(k,p) \neq 0$.  (N.B.\ $\tau_8$ herein corresponds to $\tau_4$ in the notation of Refs.\,\cite{Chang:2010hb,Chang:2011ei}.)
This was confirmed in Ref.\,\cite{Bashir:2011dp}, wherein the analysis yields a vertex \emph{Ansatz} that includes $\tau_8 = a_8 \Delta_A(k^2,p^2)$, whose functional form is precisely the same as that predicted herein, Eq.\,\eqref{eqtau8}.  We find $a_8 = -1$.  The asymptotic analysis in Ref.\,\cite{Bashir:2011dp} indicates that $1 + a_2 + 2(a_3 - a_6 + a_8) = 0$, where $a_2$ is associated with the $\tau_2$ term.  If $\{X_i\equiv 0,i=1,\ldots,8\}$, then $(a_2=0,a_3=1/2,a_6=0)$ and hence the solution to Eqs.\,\eqref{eqWGTI}, \eqref{eqtransWTI1}, \eqref{eqtransWTI2} is consistent with the known constraint.  Corrections to Eq.\,\eqref{eqtau8} involve $\{X_i,i=2,3,8\}$.  They will depend on the gauge parameter and can modify the coefficient in Eq.\,\eqref{eqtau8} on that domain within which it is meaningful to describe the vertex \emph{Ansatz} in this way.

The preceding considerations lead us to a minimal \emph{Ansatz} for the vertex that describes the interaction between an Abelian gauge boson and a dressed fermion:
\begin{equation}
\label{eqGammaminimal}
\Gamma^{\rm M}_\mu(k,p) = \Gamma_\mu^{\rm BC}(k,p) + \Gamma_\mu^{\rm TM}(k,p)\,,
\end{equation}
where $\Gamma_\mu^{\rm BC}(k,p)$ is constructed from Eqs.\,\eqref{eqGammaL}, \eqref{eqlambdaBC}, \eqref{eqlambdaG} and $\Gamma_\mu^{\rm TM}(k,p)$ is built from Eqs.\,\eqref{eqGammaT}, \eqref{eqtau}, \eqref{TsEuclidean} plus the results $\{\tau_j\equiv 0,j=1,2,4,6,7\}$.  We describe the \emph{Ansatz} as minimal because it is the simplest structure that is simultaneously compatible with the constraints expressed in Ref.\,\cite{Bashir:2011dp} and all known Ward-Green-Takahashi identities, both longitudinal and transverse.

Employed to express the electromagnetic coupling of a dressed-fermion described by a spinor that satisfies
\begin{equation}
\bar u(p,\mathpzc{M}) (i\gamma\cdot p + \mathpzc{M} ) = 0 = (i\gamma\cdot p + \mathpzc{M} )u(p,\mathpzc{M})\,,
\end{equation}
the vertex produces a renormalisation-point-invariant anomalous magnetic moment \cite{Bashir:2011dp}
\begin{equation}
\kappa = 2 \mathpzc{M} \frac{2 \mathpzc{M} \delta_A - 2 \delta_B}{\sigma_A - 2 \mathpzc{M}^2 \delta_A+2 \mathpzc{M} \delta_B}
= \frac{- 2 M \delta_{M}}{1+2 M \delta_M},
\end{equation}
where $\sigma_A = \Sigma_A(\mathpzc{M}^2,\mathpzc{M}^2)$, $\delta_{A,B,M} = \Delta_{A,B,M}(\mathpzc{M}^2,\mathpzc{M}^2)$.  In the chiral limit and absent DCSB, $\mathpzc M=0$ and hence $\kappa$ vanishes.
In contrast, using the DCSB-improved gap equation kernel in Ref.\,\cite{Chang:2013pq}, which yields a Euclidean constituent-quark mass $\mathpzc{M}=0.38\,$GeV, we find $\kappa = 1.14$.  The anomalous moment is positive, as it must be for an Abelian interaction, and commensurate with the value computed using the \emph{Ansatz} in Ref.\,\cite{Bashir:2011dp}; viz., $\kappa = 1.6$.
The computed value of $\kappa$ is large but, like the Euclidean constituent-quark mass, this is just one (infrared) value on a curve that describes the anomalous magnetic moment distribution of a dressed-quark \cite{Chang:2010hb}: averaged over a nonperturbative domain $p^2\in[0,2]\,$GeV$^2$, $\bar \kappa = 0.49$.

As with all nonperturbative quantities in QCD, the dynamical component of the $\kappa$ distribution vanishes as a power-law for $p^2\gtrsim(5\Lambda_{\rm QCD})^2$.  Thus, and for example, the dynamically generated anomalous magnetic moment influences the long-wavelength behaviour of electromagnetic form factors but is not discernible on the domain within which perturbative analyses are valid \cite{Chang:2011tx}.

\smallskip

\hspace*{-\parindent}\textbf{Epilogue}.
We elucidated novel consequences of Lagrangian-based symmetries for the fermion--gauge-boson vertex, therewith confirming numerous features of the \emph{Ansatz} described in Ref.\,\cite{Bashir:2011dp} and thus placing it both in a broader context and on firmer ground.  Our proposal, Eq.\,\eqref{eqGammaminimal}, is simpler, however.  It is thus easier to use, e.g., in Poincar\'e-covariant symmetry-preserving studies of hadron electromagnetic form factors of the type described in Refs.\,\cite{Cloet:2008re,Eichmann:2011ej}; and might serve readily as a prototype in the construction of electromagnetic currents for use in few-nucleon physics \cite{Gross:1987bu}.

Well-motivated, realistic \emph{Ans\"atze} for the dressed-quark-gluon vertex are also needed because this vertex is a critical part of all gap and Bethe-Salpeter equation studies of hadron spectra and interactions, and yet information available from continuum or lattice methods is limited \cite{Chang:2010hb,Chang:2011ei,Skullerud:2003qu,Bhagwat:2004kj,Skullerud:2004gp}.  In this connection, Eq.\,\eqref{eqGammaminimal} may be compared with the vertex in Ref.\,\cite{Chang:2011ei}, which is associated with the most realistic Poincar\'e-covariant, continuum description of the light-quark meson spectrum that is currently available.  Whilst the $\tau_{5,8}$ structures, crucial to the expression of DCSB in the spectrum, are included therein, the $\tau_{3}$ term is omitted.  This suggests both: that the \emph{Ansatz} in Ref.\,\cite{Chang:2011ei} can be improved; and a simple way by which that may be accomplished.  This phenomenological extension of the model-independent results described herein is underway.


In another direction, given that the Abelian transverse Ward-Green-Takahashi identities may now be judged useful, it is worth revisiting their non-Abelian analogues \cite{He:2009sj}, in the hope that from them our methods might distill qualitative and perhaps even semi-quantitative constraints on the dressed-quark-gluon vertex.  Realising dynamical chiral symmetry breaking in that vertex is vital to a continuum explanation of the hadron spectrum \cite{Qin:2011xq} and, therefore, to reliable predictions regarding the existence and properties of exotic mesons, that new state of matter which is conceivable in QCD but whose being is impossible in a quantum mechanics based solely on a constituent-quark, constituent-antiquark picture of mesons.  It has implications, too, for studies of QCD in-medium.  The new structures related to anomalous magnetic moments will certainly affect the location, and possibly even the existence, of a critical endpoint in the temperature-chemical potential plane \cite{Qin:2010nq}; and, given that strong fields are generated at the core of the fireball produced in a relativistic heavy ion collision, nonperturbatively generated features of the quark-gluon vertex may affect the phase transition in as yet unknown ways.

\smallskip

\setcounter{equation}{0}
\renewcommand{\theequation}{A\arabic{equation}}

\noindent\textbf{Appendix}.  Here we list the tensors used in Eqs.\,\eqref{eqGammaL}, \eqref{eqGammaT}:
\begin{equation}
\label{eqlambdaG}
\begin{array}{ll}
L^1_\mu(k,p) = \gamma_\mu \,, &
L^2_\mu(k,p) = \frac{1}{2}\, t_\mu\, \gamma\cdot t\,,\\
L^3_\mu(k,p) = - i t_\mu\, \mathbf{I}_{\rm D}\,,&
L^4_\mu(k,p) = - \sigma_{\mu\nu}\, t_\nu\,,
\end{array}
\end{equation}
%
where $\mathbf{I}_{\rm D}$ is the $4\times 4$ identity matrix in Dirac space; and
\begin{subequations}
\label{TsEuclidean}
{\allowdisplaybreaks
\begin{eqnarray}
T^1_{\mu} (k,p) &=& i \left[  p_\mu (k\cdot q) - k_\mu (p\cdot q) \right], \\
T^2_{\mu} (k,p) &=& -i T^1_{\mu} (\gamma \cdot k + \gamma \cdot p)\,,\\
\label{T3mu}
T^3_{\mu} (k,p)  &=&  q^2 \gamma_\mu - q_\mu \gamma \cdot q =: q^2 \gamma_\mu^{\rm T}\,, \\
T^4_{\mu} (k,p)  &=&   iT^1_{\mu} p_\nu k_\rho \sigma_{\nu\rho} \,,\\
T^5_{\mu} (k,p)  &=&  \sigma_{\mu\nu} q_\nu \,,\\
T^6_{\mu} (k,p)  &=&  -\gamma_\mu (k^2- p^2) + (k+p)_\mu \gamma \cdot  q \,, \\
\nonumber T^7_{\mu} (k,p)  &=&  \frac{i}{2} (k^2 - p^2) [ \gamma_\mu (\gamma \cdot  k + \gamma \cdot p) -(k+p)_\mu]\\
&& + (k+p)_\mu   p_\nu k_\rho \  \sigma _{\nu\rho} \,,\\
T^8_{\mu} (k,p)  &=&
k_\mu \gamma \cdot p
- p_\mu \gamma \cdot  k
-i \gamma_\mu p_\nu k_\rho \sigma _{\nu\rho} \,.
\end{eqnarray}}
\end{subequations}

\smallskip

%
\noindent\textbf{Acknowledgments}.
%
We thank A.~Bashir, C.~Chen, L.-j.~Jiang, M.~Pitschmann, J.~Segovia and X.-y.~Xin for valuable comments; and S.-x.~Qin is grateful for encouragement from D.\,H.~Rischke.
Work supported by:
Alexander von Humboldt Foundation Postdoctoral Research Fellowship; 
Forschungszentrum J\"ulich GmbH;
National Natural Science Foundation of China, contract nos.\ 10935001, 11075052 and 11175004;
National Key Basic Research Program of China, contract no.\ 2013CB834400;
and
U.\,S.\ Department of Energy, Office of Nuclear Physics, contract no.~DE-AC02-06CH11357.
%



\end{document}